\documentclass[12pt]{article}
\usepackage{amsmath,amssymb}
\usepackage{graphicx,color}
\numberwithin{equation}{section}
\usepackage{cite}
\usepackage{bm}
\usepackage{yfonts}
\usepackage{dcolumn}
\newcommand{\be}{\begin{equation}}
\newcommand{\ee}{\end{equation}}
\newcommand{\bea}{\begin{eqnarray}}
\newcommand{\eea}{\end{eqnarray}}
\newcommand{\beqar}{\begin{eqnarray*}}
\newcommand{\eeqar}{\end{eqnarray*}}
\newcommand{\wn}{\textswab{w}}

\renewcommand{\r}{\rho}

\newcommand{\om}{\omega}

\begin{document}
\begin{titlepage}
\hfill
\vbox{
    \halign{#\hfil         \cr
         IPM/P-2009/002  \cr
          } 
      }  
\vspace*{20mm}
\begin{center}
{\Large {\bf Non-relativistic D3-brane in the presence of higher derivative corrections }\\ }

\vspace*{15mm}
\vspace*{1mm}
{ Ahmad Ghodsi$^{a,b}$ and  Mohsen Alishahiha$^b$ }
\vspace*{1cm}

%

{\it $^a$Department of Physics, Ferdowsi University of Mashhad, \\ 
P.O.Box 91775-1436, Mashhad, Iran}\\

\vspace*{.4cm}

{\it ${}^b$School of physics, Institute for Research in Fundamental Sciences (IPM)\\
P.O. Box 19395-5531, Tehran, Iran \\ }

\vskip 2.6 cm

\end{center}

\begin{abstract}
\end{abstract}
Using ${\alpha'}^3$ terms of type IIB supergravity action we study higher order corrections
to the non-relativistic non-extremal D3-brane. Utilizing the corrected solution 
we evaluate corrections to temperature, entropy and shear viscosity. We also compute the
${\eta}/s$ ratio which  although within the range of validity of the supergravity approximation 
and in the lowest order of the correction the universal bound is respected, there is a possibility 
for a violation of the bound when higher terms in the expansion are taken into account.

\end{titlepage}

\section{Introduction}

AdS/CFT correspondence \cite{Maldacena:1997re} has been found to be a powerful tool for studying 
strongly coupled field theories in terms of  weakly coupled gravities. For instance it has provided us a framework 
to study thermodynamic and hydrodynamic properties of certain strongly coupled gauge theories which have
gravity dual (see for example \cite{Son:2007vk}). In particular in this context the AdS/CFT correspondence 
has been used to evaluate the ratio of shear viscosity to the entropy density (see for example \cite{Kovtun:2003wp,Buchel:2003tz,Buchel:2004di,
Buchel:2004qq,Benincasa:2006fu,Mateos:2006yd,Landsteiner:2007bd,Iqbal:2008by,
Buchbinder:2008nf,Garousi:2008ai}).

For example, for ${\cal N}=4$ supersymmetric Yang-Mills theory in 4D at finite temperature whose gravity 
dual is given in terms of the near horizon of the non-extremal D3-brane it was shown in 
\cite{Policastro:2001yc} that at large t'Hooft coupling and for large number of colors (large $N$) the gravity calculations give $\frac{\eta}{s}= \frac{1}{4\pi}$, 
which is compatible with the RHIC data \cite{Luzum:2008cw}. In fact motivated by gravity calculations 
it was proposed \cite{Kovtun:2003wp} that the $\eta/s$ is bounded from below
\be
\frac{\eta}{s}\geq \frac{1}{4\pi}\,.
\ee
Further investigations have confirmed this, rather universal, behavior for broader examples \cite{Myers:2008me}.
We note, however, that since in the gravity side we have higher derivative corrections to the tree level action
which has been used  
to reach to the above conclusion, one may wonder if these corrections would spoil the bound. Indeed this
point has been studied in several papers including \cite{Kovtun:2004de,Myers:2008yi,
Buchel:2008vz,Buchel:2008sh}) where it was shown that there is a possibility for the bound to be violated 
when higher order corrections are taken in to account.

In this article we would like to extend the above considerations for those non-relativistic field theories
whose gravity duals are given in 
  \cite{{Balasubramanian:2008dm},{Son:2008ye}}. The corresponding gravity solutions may be embedded in ten
dimensional type II supergravities. Indeed starting from brane solutions in type II supergravities and 
using a Null Melvin Twist \cite{{Gimon:2003xk},{Alishahiha:2003ru}} one can find new solutions
(non-relativistic branes) which  provide 
gravity duals for non-relativistic field theories \cite{{Yamada:2008if},{Mazzucato:2008tr},{Hartnoll:2008rs}}.
Following the relativistic case one would expect that heating up the non-relativistic field theory
corresponds to adding a black hole solution to the bulk \cite{{Herzog:2008wg},{Maldacena:2008wh},{Adams:2008wt},
{Schvellinger:2008bf}}. 

Having had the gravity description of the non-relativistic field theory one may proceed to study 
thermodynamic and hydrodynamic properties of the theory. 
Indeed at leading order this has been done
in \cite{{Chen:2008ad},{Kovtun:2008qy}} where it was shown that the ratio of the shear viscosity to
entropy density is the same as that in the relativistic case; $\eta/s=1/4\pi$. It is then natural to pose the question
whether there is a lower bound for this ratio in the non-relativistic case as well. 
To explore this question we will study the effects of higher order corrections to this ratio in the 
context of AdS/NRCFT.

To be specific we will consider non-relativistic non-extremal D3-brane in type IIB string theory which can be 
used to study thermodynamic and hydrodynamic properties of the dual three dimensional non-relativistic 
field theory.
In leading order the gravity solution can be obtained from D3-brane solution by making use of the Null Melvin Twist.
The resultant background
is still a solution to the leading order action of type IIB supergravity. 
Now the task is to find the corrections to the solution when ${\alpha'}^3$ terms are added to the action.
This can be used to study the effects of higher order terms to the 
$\eta/s$ ratio \footnote{Higher order corrections to the non-relativistic background have recently been 
studied in \cite{Adams:2008zk} where the higher derivative terms were given by Gauss-Bonnet action.
It was then shown that the higher order terms may correct the power exponent and as a result the
$\eta/s$ bound may be violated.}.

We note, however, that in general it is difficult to find a closed form for the corrected
solution. Nevertheless one can solve the equations of motion in the presence of higher derivative terms 
perturbatively. In fact there are two parameters that control the perturbation. The first one, denoted
by $\gamma$, controls the higher order terms in the action. The other, denoted by $\lambda$, 
parametrizes the deviation of the geometry from that in the relativistic case\footnote{To be precise the relevant
expansion parameter is $\lambda^2 T$ where $T$ is the temperature.}. Taking these two as 
the expansion parameters one finds the following schematic expansion for a typical quantity ${\cal Q}$
\be
{\cal Q}={\cal Q}_0\bigg[1+\bigg(c_1^0+c_1^1\;(\lambda^2 T)+c_1^2\;(\lambda^2 T)^2+\cdots\bigg)\gamma+\cdots\bigg],
\ee
where $c_i^j$ are some numerical factors. In particular for $\eta/s$ ratio at first subleading order 
we find
\be
\frac{\eta}{s}=\frac{1}{4\pi}\bigg[1+\bigg(120-183.31\pi^2\; (\lambda^2 T)^2+{\cal O}((\lambda^2 T)^4)\bigg)\;\gamma\bigg],
\ee
which has a potential to violate the bound. We will back to this point later.

The paper is organized as follows. In the next section we will review the tree level non-relativistic
non-extremal D3-brane to fix our notation. Then we will find higher order corrections to the solution
when the ${\alpha'}^3$ terms are added. In section three we will study the thermodynamic and 
hydrodynamic properties of the dual non-relativistic theory in the presence of higher derivative terms.
In particular we will find the corrections to the  temperature, entropy and shear viscosity. The last section 
is devoted to conclusions where we also give comments on the $\eta/s$ ratio.

\section{Higher derivative corrections}

In this section we will study higher derivative corrections to the non-relativistic non-extremal D3-brane.
To proceed we will first fix our notation by reviewing the tree level action of type IIB supergravity 
and its non-extremal D3-brane solution. When only 
dilaton, metric, B-field and RR-five form are non-zero the tree level action of type IIB 
supergravity in the string frame is given by
\bea
I_0&=&\frac{1}{2\kappa^2_{10}}\int d^{10}x\sqrt{-g}{\cal L}_0\cr &=&\frac{1}{2\kappa^2_{10}}\int d^{10}x\sqrt{-g}\bigg\{e^{-2\Phi}\left(R+4\partial_\mu\Phi\partial^\mu\Phi-\frac{1}{2\cdot 3!}H_{(3)}^2\right)
-\frac{1}{4\cdot5!}F_{(5)}^2\bigg\}\,.
\eea
The near horizon limit of the non-extremal D3-brane solution may be written as follows
\bea\label{nonD3}
ds^2&=&r^2\left(-f_0^2(r) dt^2+dy^2+dx_1^2+dx_2^2\right)+
\frac{dr^2}{f_0^2(r) r^2}+(d\chi+A)^2+ds_{CP^2}^2\,,\quad  e^{-2\Phi}=1\,,\cr &&\cr
F_{(5)}&=&dC_{(4)}=2(1+*)d\chi \wedge J \wedge J\,,\quad dA=2J\,,\quad Vol(CP^2)=\frac{1}{2} J \wedge J\,. 
\eea
Here $f_0^2(r)=1-r_0^4/r^4$ and
\bea
A=\frac{1}{2}\sin^2\mu(d\psi+\cos\theta d\phi)\,,\,\,\,\quad
ds^2_{CP^2}=d\mu^2+\frac14\sin^2\mu(\sigma_1^2+\sigma_2^2+\cos^2\mu\sigma_3^2)\,,
\eea
where $\sigma_i$ are the $SU(2)$ left invariant one-form
\bea
\sigma_1=\cos\psi d\theta+\sin\theta\sin\psi d\phi\,,\quad
\sigma_2=-\sin\psi d\theta+\sin\theta\cos\psi d\phi\,,\quad
\sigma_3=d\psi+\cos\theta d\phi\,.
\eea

Using the Null Melvin Twist one can map the solution \eqref{nonD3} to the non-relativistic non-extremal D3-brane 
which is still a solution of the tree level type IIB supergravity action. 
The obtained solution is \cite{Mazzucato:2008tr}
\bea
ds^2&=&\frac{r^2f_0^2(r)}{k(r)}\left(-(1+r^2\lambda^2)dt^2+\frac{1-r^2\lambda^2 f_0^2(r)}{f_0^2(r)}dy^2
-2r^2\lambda^2dtdy\right)\cr &&\cr
&+&r^2(dx_1^2+dx_2^2)+\frac{dr^2}{f_0^2(r)r^2}+\frac{1}{k(r)}(d\chi+A)^2+ ds_{CP^2}^2\,,\cr &&\cr
B_{(2)}&=&\frac{r^2\lambda }{k(r)}(d\chi+A)\wedge(f_0(r)dt+dy)\,,\quad e^{2\Phi}=\frac{1}{k(r)}\,, 
\eea
where $k(r)={1+r^2\lambda^2(1-f_0^2(r))}$ and the RR-five form remains unchanged.

Now the aim is to study the effects of higher derivative terms to the non-relativistic non-extremal D3-brane solution.
Here we restrict ourselves to corrections from gravity side. There are other efforts have been done to include other correction terms, for example see \cite{Myers:2008yi}. The higher derivative terms which we are interested in are the  ${\alpha'}^3$ corrections to type IIB supergravity action. In string frame they are given by 
\bea 
I_1 =\frac{1}{2\kappa^2_{10}}\int d^{10}x\,\sqrt{-g}\;\gamma{\cal L}_1
=\frac{1}{2\kappa^2_{10}}\int d^{10}x\,\sqrt{-g}\; \gamma e^{-2\phi}W\,,
\eea 
where $\gamma=\frac18\zeta(3)(\alpha')^3$ and $W$ can be written in terms 
of the Weyl tensors
\be
W=C^{hmnk}C_{pmnq}{C_h}^{rsp}{C^q}_{rsk}+\frac12 C^{hkmn}C_{pqmn}{C_h}^{rsp}{C^q}_{rsk}\,.
\ee

In principle one needs to solve the equations of motion coming from the action $I_0+I_1$.
Of course, in general, it is not an easy task to solve the resultant equations. 
Nevertheless we may start from an ansatz and try to solve 
the equations of motion for the parameters of the ansatz. We note, however, that in general the 
obtained equations cannot be solved exactly, though they may be solved perturbatively.

For making our ansatz, we start from the same general ansatz as \cite{Pawelczyk:1998pb} for relativistic non-extremal D3-branes. Then we do all the steps for Null Melving Twisting of $D_p$-branes noted in \cite{Mazzucato:2008tr} for this general background. Finally we find the following ansatz for the non-relativistic non-extremal D3-branes 
\bea
ds^2&=&\frac{r^2f^2(r)}{k(r)}\left(-(1+r^2\lambda^2 s^2(r))dt^2+\frac{1-r^2\lambda^2 s^2(r) f^2(r)}{f^2(r)}dy^2
-2r^2\lambda^2 s^2(r)dtdy\right)\cr &&\cr
&+&r^2(dx_1^2+dx_2^2)+\frac{h^2(r)}{r^2}dr^2+\frac{s^2(r)}{k(r)}(d\chi+A)^2+s^2(r) ds_{CP^2}^2\,,\cr &&\cr
F_{(5)}&=&dC_{(4)}=2(1+*)d\chi \wedge J \wedge J\,,\quad \quad \quad e^{2\Phi}=\frac{e^{2v(r)}}{k(r)}\,,\cr &&\cr
B_{(2)}&=&\frac{r^2\lambda s^2(r)}{k(r)}(d\chi+A)\wedge(f(r)dt+dy)\,,
\eea
for arbitrary functions $f(r),h(r),s(r)$ and $v(r)$. Moreover 
\be
k(r)=1+r^2\lambda^2s^2(r)(1-f^2(r))\,.
\ee
Note that in the limit of $\lambda\rightarrow 0$ the above ansatz reduces to that in the 
relativistic case. We will also assume that $f(r)$ and $h(r)$ have zero at $r=\pm r_0$ and the
extremal limit is obtained at $r_0\rightarrow 0$.

Using the above ansatz the equations of motion of the total action $I_0+I_1$ read
\be
\frac{\partial {\cal L}_0}{\partial X_i}-\frac{d}{dr}\frac{\partial {\cal L}_0}{\partial X'_i}+\frac{d^2}
{dr^2}\frac{\partial {\cal L}_0}{\partial X''_i}=-\gamma (\frac{\partial {\cal L}_1}{\partial X_i}-
\frac{d}{dr}\frac{\partial {\cal L}_1}{\partial X'_i}+\frac{d^2}{dr^2}\frac{\partial {\cal L}_1}{\partial X''_i})\,,
\ee
where prime denotes the derivative with respect to $r$ and $X_i=\{f(r),h(r),s(r),v(r)\}$. To proceed we 
choose the following perturbed ansatz over the tree level solution
\bea
f(r)&=&{(1-\frac{r_0^4}{r^4})^\frac12}\left(1+\gamma F(r)\right)\,, \quad
h(r)={(1-\frac{r_0^4}{r^4})^{-\frac{1}{2}}}\left(1+\gamma H(r)\right)\,,\cr &&\cr
s(r)&=&1+\gamma S(r)\,,\qquad \qquad \qquad \, \ v(r)=\gamma V(r)\,.
\eea
It is straightforward, though messy, to plug the ansatz to the equations of motion to find the
following differential equations for the perturbed functions $\{F(r),H(r),S(r),V(r)\}$. Setting 
$\rho=\frac{r}{r_0}$ and $\tilde{\lambda}=\lambda r_0$ one gets
\bea\label{eom1}
& &(\rho^4-1)\, (\rho\, S''-\frac25 \rho\, V''-\frac35 H')+4\,(\rho^4-\frac12)\,(S'
-\frac25 V')-4 \rho^{3}\,(S-\frac25 V+\frac{3}{5} H)\cr &&\cr
&=&\frac{F_1(\rho;\tilde{\lambda})}{\rho^{13}(\rho^2+\tilde{\lambda}^2)^9}\,,
\eea
\bea\label{eom2}
& &\rho\,(\rho^4-1)\,(F''+4S''-2V'')+10(\rho^4-\frac15 )\,(S'-2 V')-4\,(\rho^4-\frac12 )\,H'\cr &&\cr
&+&\rho^{3}\,(6\rho\, F'+52 S-8 V-20 H)=\frac{F_2(\rho;\tilde{\lambda})}{\rho^{13}(\rho^2+\tilde{\lambda}^2)^9}\,,
\eea
\bea\label{eom3}
(\rho^4-1)\,F'+\frac{4}{3}(\rho^4-\frac12)\,( 5 S'-2 V' )
-4\rho^{3}\,(H+\frac{5}{3}S-\frac23 V)=
\frac{F_3(\rho;\tilde{\lambda})}{\rho^{13}(\rho^2+\tilde{\lambda}^2)^8}\,,
\eea
\bea\label{eom4}
& &{\rho}\, (\rho^4-1)\,(F''+5S''-2V'')+5({\rho}^{4}-\frac15)\,(5 S'-2 V')
-4({\rho}^{4}-\frac12)\,H'\cr &&\cr
&+&{\rho}^{3}\,(6\rho F'-20 H+20S) =\frac{F_4(\rho;\tilde{\lambda})}{\rho^{13}(\rho^2+\tilde{\lambda}^2)^8}\,.
\eea
Here $F_i$ are functions of $\rho$ whose explicit forms are given in appendix A. Now the task is to 
perturbatively solve the above equations with given boundary conditions. 
First of all we would like to have a solution whose asymptotic is the same as that for the
non-relativistic non-extremal D3-brane. Moreover we require that the solution to be finite at the horizon.

For our purpose we just need to solve the equations up to ${\cal O}(\tilde{\lambda}^2).$\footnote{
Note that using the expression of the temperature, \eqref{Tem}, at leading order one has
$\tilde{\lambda}=\lambda r_0=\lambda^2 T$.} The results are
\bea
F(\rho)\!\!&=&\!\!{\frac {640-1375\rho^{4}-1465{\rho}^{8}}{32 \rho^{12}}}+ \bigg( \frac{11978}{15}( \rho^4-{\frac {2837}{5989}})\ln( 1+\frac{1}{\rho^{2}})-\frac{6304}{15}\ln(2) \cr &&\cr
\!\!&+&\!\!\frac{1564664}{1575}-\frac{11978}{15}{\rho^{2}} +\frac{94}{45}\frac{1}{\rho^{2}}+\frac{2197}{450}\frac{1}{\rho^{6}}
-\frac{109523}{350}\frac{1}{\rho^{10}}+\frac{2332}{21}\frac{1}{\rho^{14}} \bigg) \frac {\tilde{\lambda}^{2}}{\rho^{4}-1} \,.
\eea

\bea
H(\rho)&=&\frac {2480\rho^8+1895\rho^4-10000}{64 \rho^{12}}+\bigg(( \frac{5989}{5}{\rho^{8}}-
\frac{9036}{5}{\rho^4}+\frac{2837}{15})\ln( 1+\frac{1}{\rho^{2}} )\cr &&\cr
&+&\frac{6304}{15}\ln(2)-\frac{1657358}{1575}-\frac{18777}{28}\frac{1}{\rho^{14}}+\frac{45493}{75}\frac{1}{\rho^{10}}+\frac{607097}{6300}\frac{1}{\rho^6}+\frac{24803}{75}\frac{1}{\rho^2}\cr &&\cr
&+&\frac{21119}{15}{\rho^2}+\frac{11978}{25}{\rho^4}-\frac{5989}{5}{\rho^6}\bigg)
\frac {\tilde{\lambda}^{2}}{\rho^{4}-1}\,.
\eea

\bea
S(\rho)&=&-{\frac {15+90\rho^{4}}{64 \rho^{8}}}+ 
\bigg(({\frac {5989}{25}}{ {\rho^{4}}}-{\frac {3257}{25}})\ln(1+\frac{1}{\rho^2})-{\frac {5989}{25}}\rho^{2}+{\frac {3782}{75}}{\frac {1}{\rho^{2}}}
-{\frac {1532}{375}}{\frac {1}{\rho^{6}}}\cr &&\cr
&+&{\frac {184273}{10500}}{\frac {1}{{\rho}^{10}}}-{\frac {53}{20}}{\frac {1}{\rho^{14}}} \bigg) \tilde{\lambda}^{2}\,.
\eea

\bea
V(\rho)&=&-{\frac {30+45\rho^{4}+90\rho^{8}}{16 \rho^{12}}}
-{\frac {1}{175}} \bigg( 83846 +7350\ln  ( 1+\frac{1}{\rho^2} )-7350\frac{1}{\rho^{2}}-2450\frac{1}{\rho^{6}} \cr &&\cr
&-&16170\frac{1}{\rho^{10}}+3375\frac{1}{\rho^{14}}\bigg) 
\tilde{\lambda}^{2}\,.
\eea
These are all information we need to evaluate the corrections to the temperature, entropy and shear viscosity
which we will do in the next section. It is worth noting that the above solution,
after going to Einstein frame and using a field redefinition, reduces to the known results 
in \cite{Pawelczyk:1998pb} for $\tilde\lambda=0$ .


\section{Entropy and Viscosity}

In this section we will study thermodynamic and  hydrodynamic properties of the finite temperature non-relativistic three dimensional field theory whose gravity dual is given in terms of the non-relativistic non-extremal D3-brane.
In fact, by making use of the results of the previous section, we will be able to study the effects of the 
higher derivative corrections to the entropy and the shear viscosity. 
To proceed let us start from the thermodynamical quantities.

\subsection*{Temperature}
Since in the presence of the higher derivative terms  the solution gets correction, one expects that 
the temperature will also be corrected. To find the temperature one may follow the standard route. Namely we will have to identify the Euclidean time in the corrected geometry in such a way that 
the corrected metric will be free of conical singularity. Consider $\zeta^\mu$ as the null generator of the horizon, the surface gravity is defined by
\be
\kappa^2=-\frac{1}{2}\nabla^\mu\zeta^\nu\nabla_\mu\zeta_\nu\,,
\ee
and the temperature is related to surface gravity by $T=\frac{\kappa}{2\pi}$. As indicated in \cite{Herzog:2008wg}
the Killing generator of the horizon is given by $\zeta=\frac{1}{\lambda}\frac{\partial}{\partial t}$ which differs from the relativistic case by a prefactor. Knowing these, to first order of $\gamma$ and $\tilde{\lambda}^2$ one finds
\bea\label{Tem}
 T&=&\left[(1+\frac{1225}{64}\gamma)+(\frac{3152}{15}\ln(2)-\frac{206806}{1575})\gamma \tilde{\lambda}^2\right]\frac{r_0}{\pi\lambda}\cr &&\cr
 &=&\left[1+(19.14 +14.35 \tilde{\lambda}^2)\gamma\right] \frac{r_0}{\pi\lambda}\,.
\eea  
Note that in the limit of $\gamma\rightarrow 0$ this reduces to known leading order expression \cite{Herzog:2008wg}. On the other hand
for $\tilde{\lambda}\rightarrow0$ and $\lambda T\rightarrow T$ we recover the relativistic result as well \cite{Pawelczyk:1998pb,Buchel:2004di}.

\subsection*{Entropy}

Entropy is one of the interesting quantities usually people study in this context. To 
compute the entropy one may use the Euclidean method by which we need to evaluate the value of the action on 
our solution. We note, however, that in general the action diverges and needs to be regularized by 
subtracting the value of the action evaluated on a reference background. To do this we will have to be careful
about how to define the cut off. Although this method is found useful for geometries which are 
asymptotically $AdS$, in general for geometries with arbitrary asymptotic, this procedure is not easy to be
applied. Therefore it will be useful if one can use another method.

Indeed there is an alternative way to compute the entropy via Wald's formula for the 
entropy \cite{Wald:1993nt,{Jacobson:1993vj},{Iyer:1994ys}}. When the Lagrangian does not contain the covariant derivative of the curvature
the Wald formula for the entropy reduces to the following simple form 
\cite{Visser:1993nu}
\be\label{Wald}
S_{BH}=\int_H dx^H\sqrt{g^H}\frac{\partial {\cal L}}{\partial R_{\mu\nu\lambda\rho}}g^{\bot}_{\mu\lambda}\,g^{\bot}_{\nu\rho}\,,
\ee
where ${\cal L}$ is the total Lagrangian and the integration is over the horizon. $g^{\bot}$ is the 
orthogonal metric to the horizon which can be obtained in terms of the normal vectors to the horizon 
as follows
\bea
g_{\mu\nu}^{\bot}=(N_t)_\mu(N_t)_\nu+(N_r)_\mu(N_r)_\nu\,.
\eea
In our case setting ${\mathcal{L}}=\frac{4\pi}{2\kappa^2_{10}}({\cal L}_0+\gamma {\cal L}_1)$ and taking 
into account that 
\bea
N_t=\sqrt{g_{tt}}(1,\frac{g_{ty}}{g_{tt}},0,0,0,\vec{0})\,,\quad N_r=(0,0,0,0,\sqrt{g_{rr}},\vec{0})\,,
\eea
the Wald entropy \eqref{Wald} reads (for more technical details see e.g. \cite{Ghodsi:2006cd})
\bea
S_{BH}=\frac{8\pi}{2\kappa^2_{10}}\int_H dx^H\sqrt{g^H}\bigg\{\frac{\partial {\mathcal{L}}}{\partial  R_{trtr}}g^{\bot}_{tt}\,g^{\bot}_{rr}+\frac{\partial {\mathcal{L}}}{\partial  R_{yryr}}g^{\bot}_{yy}\,g^{\bot}_{rr}+2\frac{\partial {\mathcal{L}}}{\partial  R_{tryr}}g^{\bot}_{ty}\,g^{\bot}_{rr}\bigg\}\,.
\eea
Using the corrected geometry presented in the previous section it is easy to compute each term in the 
above expression for the entropy. Altogether up to ${\cal O}(\tilde{\lambda}^2)$ one arrives at
\bea
S_{BH}&=&\frac{2\pi^2RV_2V_{CP^2}V_1}{2\kappa^2_{10}}r_0^3\bigg((4+\frac{4635}{16}\gamma)+(\frac{12608}{5}\ln(2)-\frac{1356424}{525})\gamma \tilde{\lambda}^2\bigg)\,,\cr &&\cr
&=&\frac{2\pi^2RV_2V_{CP^2}V_1}{2\kappa^2_{10}}r_0^3\bigg((4+289.68\;\gamma)-835.82\;
\gamma\;\tilde{\lambda}^2\bigg)\,.
\eea
Here $R$ is the radius of compact direction $y$, $V_1$ is the volume of the $\chi$ coordinate and $V_2$ is the volume of $x_1$ and $x_2$ directions. Using the expression for the temperature \eqref{Tem}
we can eliminate $r_0$ to express the entropy in terms of the temperature 
\be
S_{BH}=\frac{8\pi^5RV_2V_{CP^2}V_1}{2\kappa^2_{10}}(\lambda T)^3\bigg[1+\bigg(15-252\pi^2
 (\lambda^2 T)^2\bigg)\gamma \bigg]\,.
\ee

\subsection*{Shear viscosity}

The shear viscosity of a translation invariant theory at the hydrodynamic limit can be evaluated by making use 
of the Kubo formula \cite{Son:2002sd}
\bea
\eta=-\lim_{\omega\rightarrow 0} \frac{1}{\omega} \Im G^{R}_{x_1x_2,x_1x_2}(\omega,k=0)\,,
\eea
where $G^{R}_{x_1x_2,x_1x_2}$ is the retarded Green's function of the component $T_{x_1x_2}$ of the
energy momentum tensor. To compute the retarded Green's function we utilize the AdS/CFT dictionary \cite{Son:2002sd,Herzog:2002pc}, by which
the retarded Green's function can be related to the on-shell bulk action 
\bea
G^R_{x_1x_2,x_1x_2}(\omega,{k})=\lim_{r\rightarrow \infty}2{\cal F}(\omega,{k},r)\,.
\eea

The shear viscosity in the non-relativistic field theory has been evaluated at leading order in 
\cite{Kovtun:2008qy} where it was shown that the ratio of the shear viscosity to the entropy density 
saturates the universal bound
\be
\frac{\eta}{s}=\frac{1}{4\pi}\,.
\ee
Following \cite{Buchel:2004di} we would like to study the effects of higher derivative corrections to the value of the shear viscosity.  In order 
to evaluate the shear viscosity we consider a perturbation to the metric as 
$h_{x_1x_2}=r^2 \varphi(t,r,x_1,x_2)$ which can be Fourier transformed along $t$ and $x_i$'s directions
leading to
\bea
\varphi(t,r,y,x_1,x_2)=\int \frac{d\omega d^3k}{(2\pi)^4}e^{-i\omega t+ ik\cdot\mathbf{x}}\varphi_{\omega, k}(r)\,. 
\eea 
Since in the Kubo formula, the spatial
momentum is zero, we restrict our calculations to $k=0$. On the other hand since the energy 
momentum tensor has zero particle number the perturbation must be $y$-independent.
Actually it is easy to see that 
the relevant perturbation of the metric decouples from other perturbations and in fact it obeys the equation of 
motion of a minimally coupled massless scalar \cite{Kovtun:2004de} which in our case we have 
\bea\label{eomphi}
A\varphi''_\om+C\varphi'_\om+2D\varphi_\om-\frac{d}{dr}(F\varphi''_\om+2B\varphi'_\om+C\varphi_\om)+\frac{d^2}{dr^2}(2E\varphi''_\om+F\varphi'_\om+A\varphi_\om)=0\,.
\eea 
The coefficients $A,...,F$ are given in the appendix B.
We note, however, that in general it is difficult to find an explicit solution for the above equation. Nevertheless
we may solve it perturbatively. Changing the variables as before, $\rho=\frac{r}{r_0}$ and $\tilde{\lambda}=\lambda r_0$, and expanding the above equation up to ${\cal O}(\tilde{\lambda}^2)$, one
finds
\bea
& &\rho( {\rho}^{4}-1)^2 {\frac {d^{2}\varphi_\om}{d{\rho}^{2}}} + ( 5{\rho}^{8}-6{\rho}^{4}+1) {\frac {d\varphi_\om}{d\rho}} -4{\rho}^{3}{\Omega}^{2} ({\tilde{\lambda}}^{2}{\rho}^{4}-{\rho}^{2} -\tilde{\lambda}^{2}) \varphi_\om \cr &&\cr
&+& \gamma\frac{1}{{\rho}^{14} ( {\rho}^{4}-1) }
\bigg( C_1  {\frac {d^{4}\varphi_\om}{d{\rho}^{4}}}+C_2{\frac {d^{3}\varphi_\om }{d{\rho}^{3}}}+C_3{\frac {d^{2}\varphi_\om}{d{\rho}^{2}}}+C_4{\frac {d\varphi_\om}{d\rho}}+C_5 \varphi_\om\bigg)=0\,,
 \eea
where the coefficients $C_i$ are given in appendix C. Next we set
\be
\varphi_\om(\rho)=(1-\frac{1}{\rho^2})^{\Delta} G_\om(\rho)\,,
\ee
where $G_\om(\rho)$ is a regular function at horizon and $\Delta$ is a constant to be determined. Setting
$\om=2r_0 \wn$, for a solution normalized to one at $\rho\rightarrow \infty$ up to the order of 
${\cal O}(\wn^2)$ we have
\bea
\varphi_\wn(\rho)&=&
( 1-\frac{1}{\rho^2} ) ^{\Delta(\wn)}\, \bigg\{1+\gamma \frac{\tilde{\lambda}^2 }{15}\bigg( 120\ln(1+{
\frac {1}{{\rho}^{2}}} ) -\frac{120}{\rho^2}+\frac{155}{\rho^6}+\frac{67}{\rho^{10}} \bigg)\\
\!\!\!&+&\!\!\!\Delta(\wn)\,\bigg(\ln( 1+{\frac {1}{{\rho}^{2}}})-\frac{\gamma}{16} ( \frac{3385}{\rho^4}+\frac{2335}{\rho^8}+\frac{768}{\rho^{12}}) +\gamma{\tilde{\lambda}}^{2}W ( \rho )\bigg) \bigg\}.\nonumber 
\eea
Here $W(\rho)$ is an unknown function satisfying the following differential equation
\bea
& &{\rho}^{16} ( {\rho}^{4}-1 ) ^{3}{\frac {d^{2}W ( \rho )}{d{\rho}^{2}}} - ({\rho}^{15}- 7{\rho}^{19}+11{\rho}^{23}-5{\rho}^{27} ) {\frac {dW ( \rho )}{d\rho}}-{\frac {201728}{15}}{\rho}^{18}\ln  ( 1+\rho^2)\cr &&\cr 
&+&\frac13 \bigg( 96{\rho}^{24}-660{\rho}^{20}+1344{\rho}^{16}+248{\rho}^{12}-2368{\rho}^{8}
+1340{\rho}^{4}\bigg)\ln(1+\frac{1}{\rho^2})\cr &&\cr
&-&32\bigg({\rho}^{22}-29{\rho}^{20}+{\frac {7278127}{12600}}{\rho}^{18}-{\frac {9859}{15}}{\rho}^{16}-{\frac {827}{120}}{\rho}^{14}-{\frac {184637}{360}}{\rho}^{12}-{\frac {431}{120}}{\rho}^{10}\cr &&\cr &+&{\frac {12381}{200}}^{14}{\rho}^{8}+{\frac {201}{40}}{\rho}^{6}+{\frac {96023}{56}}{\rho}^{4}-{\frac {9205}{8}}-{\frac {12608}{15}}^{4}{\rho}^{18}\ln  (4\rho) \bigg)=0\,.
\eea
Using Maple one can solve the above equation which in turn can be used to find the solution $\varphi_\wn$. 
On the other hand the constant $\Delta$ can be fixed by the regularity condition on $G_\om(\rho)$.
Following \cite{Buchel:2008sh} we note that in order to have the regularity up to order $\gamma$,  we need to 
check the regularity condition up to order of $O(\wn^2)$. Doing so one finds 
\be
\Delta(\wn)=-\frac{2i r_0\wn}{ 4\pi T}=-\frac{1}{2\lambda} i \wn \bigg[1+\left(-\frac{3152}{15}\ln(2)r_0^2\lambda^2+\frac{206806}{1575}r_0^2\lambda^2-\frac{1225}{64}\right)\gamma\bigg]\,.
\ee 
Using the  equation of motion (\ref{eomphi}) one finds the quadratic part of the on-shell action of the 
bulk theory  as follows (for more detail see \cite{Buchel:2004di}) 
\bea
\mathcal{F}(\omega,0,r)&=&-\frac{V_{CP^2}V_1}{2 \kappa^2_{10}}\bigg[\left(B-A-\frac{F'}{2}+2p_0E\right)\varphi'_\om\varphi_{-\om}+\frac12(C-A')\varphi_\om\varphi_{-\om}+Ep_1\varphi'_\om\varphi'_{-\om}\cr
&&\cr
&&\;\;\;\;\;\;\;\;\;\;\;\;\;\;\;\;\;
-E'\varphi''_\om\varphi'_{-\om}+E\varphi''_\om\varphi'_{-\om}-E\varphi'''_\om\varphi_{-\om}\bigg]\,.
\eea 
Putting everything together we get 
\bea
G^R_{x_1x_2,x_1x_2}(\wn,0)&=&\lim_{\r\rightarrow \infty}\,2\mathcal{F}(\wn,0,\r)=\lim_{\r\rightarrow \infty}\,-\frac{2V_{CP^2}V_1}{2\kappa^2_{10}}r_0^4\bigg[( 2{\r}^{2}{\tilde\lambda}^{2}-1+i\wn ) +{\r}^{4}\cr &&\cr
&+&{\frac {3232}{5}} \bigg(  ( i\wn-{\frac {394}{303}} ) {\tilde\lambda}^{2}\ln  ( 2 ) -{\frac {220583}{169680}} ( -{\frac {380537}{661749}}+i\wn ) {
\tilde\lambda}^{2}\cr &&\cr
&+& ( -{\frac {2825}{51712}}+{\frac {7325}{25856}}{\r}^{2}{\tilde\lambda}^{2}+{\frac {61575}{206848}}i\wn ) 
+{\frac {5989}{8080}}{\r}^{4}{\tilde\lambda}^{2} \bigg) \gamma\bigg].
\eea
Finally the shear viscosity becomes
\bea
\eta&=&\lim_{\omega\rightarrow 0}\,\frac{2V_{CP^2}V_1\wn }{2\kappa^2_{10}\omega}r_0^4\,\bigg(1+\frac{12315}{64}\gamma+(\frac{3232}{5}\ln(2)-\frac{441166}{525})\gamma
\tilde{\lambda}^2 \bigg)
\eea
Using the expression for the temperature \eqref{Tem} we can eliminate $r_0$ to write the shear viscosity 
in terms of the temperature as follows
\bea
\eta=\frac{\pi^3   V_{CP^2}V_1}{2\kappa^2_{10}}(\lambda T)^3\bigg[1+\bigg(135-435.31\pi^2 
(\lambda^2 T)^2\bigg)\gamma\bigg]\,.
\eea

\section{Conclusions}

In this paper we have studied the non-relativistic non-extremal D3-brane in the presence of higher order 
corrections given by ${\alpha'}^3$ terms in type IIB supergravity. By making use of 
 the corrected solution we have 
computed the effects of the higher order corrections to the temperature, entropy and shear viscosity.
Having computed the corrected entropy and shear viscosity it is natural to obtain the ratio of the 
shear viscosity to entropy density. To find it, we first need the entropy density which turns out to be
\be
s=\frac{4\pi^4 V_{CP^2}V_1}{2\kappa^2_{10}}(\lambda T)^3\bigg[1+\bigg(15-252\pi^2 (\lambda^2 T)^2\bigg)\gamma \bigg]\,.
\ee
Using the expression for shear viscosity one finds
\bea
\frac{\eta}{s}&=&\frac{1}{4\pi}\bigg(1+120\gamma+\left(16\ln(2)-\frac{972}{5}\right)\pi^2\gamma 
 (\lambda^2 T)^2\bigg)\cr &&\cr
&=&\frac{1}{4\pi}\bigg[1+\bigg(120-183.31\pi^2\; (\lambda^2 T)^2\bigg)\;\gamma\bigg],
\eea
which presents the correction to the tree level bound $1/4\pi$. Note that in the limit of 
$\lambda^2 T=0$ we recover the result of the relativistic field theory \cite{Buchel:2008sh}.
When $\gamma=0$ we get $\eta/s=1/4\pi$ which is independent of $\lambda$ showing that at tree 
level as far as the ratio is concerned both relativistic and non-relativistic cases lead to 
a universal result \cite{Kovtun:2008qy}. 

It is now important to see whether this correction can violate the lower bound of the $\eta/s$ ratio.
This depends on the sign of the $\gamma$ term,  $120-183.31\pi^2\; (\lambda^2 T)^2$. In fact
for $\lambda^2 T > 0.257$ the coefficient of $\gamma$ is negative leading to a violation of the 
bound. On the other hand for $\lambda^2 T<0.257$ the higher order corrections respect the
bound. Therefore it is crucial to estimate $\lambda^2 T$ as precise as we can.
To find a bound on the value of the corrections we note that the supergravity approximation is applicable
when the curvature is small\footnote{In our notation we have set the radius of the space time 
to one.} which in our notation it means $|{\mathcal{R}}|=20 r_0^2 \lambda^2\ll 1$, in other words, 
$\lambda^2 T \ll 0.07$. Therefore within the supergravity approximation and to the  
order of ${\mathcal{O}(\lambda^4T^2)}$ the bound is not violated. On the other hand if one considers the effects of higher orders, {\it e.g.} ${\mathcal{O}(\lambda^8T^4)}$, since the corrections may decrease the
lower bound of $0.257$, there could be a possibility for violation of the bound in a narrow window 
where $ \lambda^2 T\ll 0.07$. 
It would be interesting to explore this point better.

\vspace*{1cm}

{\bf Acknowledgments}
 
We would like to thank A. Akhavan, A. Davody, R. Fareghbal, A. E. Mosaffa, O. Saremi, S. Rouhani and 
A. Vahedi for discussions on different aspects of non-relativistic field theories and their gravity duals.
The work M. A. is supported in part by Iranian TWAS chapter at ISMO.

\section*{Appendix A}
The $F_i(\rho;\tilde{\lambda})$ functions on the right hand side of equations of motion in (\ref{eom1}) to (\ref{eom4}) are given by
\bea
F_1\!\!\!&=&\!\!\!\frac{1296}{5}\bigg[\tilde{\lambda}^4\r^{34}+\frac{91273}{15552}\tilde{\lambda}^6\r^{32}
+(\frac{2003147}{93312}\tilde{\lambda}^4-\frac{101}{108})\tilde{\lambda}^4\r^{30} 
+(\frac{9179675}{279936}\tilde{\lambda}^{8}+\frac{19861}{4374}\tilde{\lambda}^4\cr &&\cr 
\!\!\!&-&\!\!\!\frac{485}{648})\tilde{\lambda}^2\r^{28}+(\frac{2742697}{139968}\tilde{\lambda}^{8}-\frac{17461043}{2239488}\tilde{\lambda}^4+\frac{8569}{1944})
\tilde{\lambda}^4\r^{26}+\frac{3091}{972}(\tilde{\lambda}^{12}-\frac{83656099}{7121664}\tilde{\lambda}^8\cr &&\cr 
\!\!\!&-&\!\!\!\frac{72409}{148368}\tilde{\lambda}^4+\frac{4827}{6182})\tilde{\lambda}^2\r^{24}+\frac{5545}{648}(\tilde{\lambda}^{16}-\frac{116087}{19962}\tilde{\lambda}^{12}+\frac{1530733}{958176}\tilde{\lambda}^8-\frac{193517}{133080}\tilde{\lambda}^4\cr &&\cr 
\!\!\!&+&\!\!\!\frac{432}{1109})\r^{22} +\frac{206}{81}(\tilde{\lambda}^{16}+\frac{132625}{59328}\tilde{\lambda}^{12}-\frac{388271}{474624}\tilde{\lambda}^8-\frac{2123}{1648}\tilde{\lambda}^4
+\frac{4075}{1648})\tilde{\lambda}^2\r^{20}-\frac{1463}{1944}(\tilde{\lambda}^{16}\cr &&\cr 
\!\!\!&-&\!\!\!\frac{1229863}{19152}\tilde{\lambda}^{12}+\frac{5417857}{93632}\tilde{\lambda}^8
-\frac{309279}{5852}\tilde{\lambda}^4+\frac{405}{77})\r^{18} -\frac{407}{432}(\tilde{\lambda}^{16}-\frac{471325}{43956}\tilde{\lambda}^{12}\cr &&\cr
\!\!\!&-&\!\!\!\frac{4332419}{1054944}\tilde{\lambda}^8-\frac{1694084}{32967}\tilde{\lambda}^4+\frac{13396}{1221})\tilde{\lambda}^2\r^{16}+\frac{12197}{5832}(\tilde{\lambda}^{12}-\frac{44747}{73182}\tilde{\lambda}^8+\frac{49288231}{1170912}\tilde{\lambda}^4\cr &&\cr
\!\!\!&-&\!\!\!\frac{1601631}{97576})\tilde{\lambda}^4\r^{14}+\frac{137}{324}(\tilde{\lambda}^{12}+\frac{122791}{13152}\tilde{\lambda}^8+\frac{44678345}{236736}\tilde{\lambda}^4-\frac{411869}{3288})\tilde{\lambda}^6\r^{12}+\frac{20891}{8748}(\tilde{\lambda}^8\cr &&\cr
\!\!\!&+&\!\!\!\frac{14499791}{668512}\tilde{\lambda}^4-\frac{141442003}{5348096})\tilde{\lambda}^8\r^{10}+\frac{15707}{34992}(\tilde{\lambda}^8+\frac{784154}{15707}\tilde{\lambda}^4-\frac{121617627}{1005248})\tilde{\lambda}^{10}\r^8\cr &&\cr
\!\!\!&+&\!\!\!\frac{409103}{69984}(\tilde{\lambda}^4-\frac{4562105}{818206})\tilde{\lambda}^{12}\r^6
+\frac{901}{1296}(\tilde{\lambda}^4-\frac{3603757}{194616})\tilde{\lambda}^{14}\r^4
-\frac{46409}{15552}\tilde{\lambda}^{16}\r^2-\frac{42715}{139968}\tilde{\lambda}^{18}\bigg].\nonumber
\eea
\bea
F_2\!\!\!&=&\!\!\!
-\frac{2876}{5} \bigg(\tilde{\lambda}^4\r^{30} -(\frac{933871}{621216} \tilde{\lambda}^4 
-\frac{10}{719} )\tilde{\lambda}^2 \r^{28}-(\frac{12450601}{4969728} \tilde{\lambda}^4 -\frac{365}{719})\tilde{\lambda}^4 \r^{26}+(\frac{57784183}{4969728} \tilde{\lambda}^8 \cr &&\cr 
\!\!\!&+&\!\!\!\frac{38395}{6471} \tilde{\lambda}^4-\frac{24265}{8628}) \tilde{\lambda}^2 \r^{24}+(\frac{5721773}{207072} \tilde{\lambda}^{8} -\frac{287905}{138048}\tilde{\lambda}^4-\frac{230515}{34512})\tilde{\lambda}^4 \r^{22} 
+ \frac{108895}{5752} (\tilde{\lambda}^{12}\cr &&\cr 
\!\!\!&-&\!\!\!\frac{10478339}{23521320} \tilde{\lambda}^8 
-\frac{775829}{980055} \tilde{\lambda}^4 +\frac{61664}{326685}) \tilde{\lambda}^2 \r^{20}+\frac{10465}{1438}  ( \tilde{\lambda}^{16}-\frac{582367}{347760} \tilde{\lambda}^{12} +\frac{7134823}{6027840} \tilde{\lambda}^8 \cr &&\cr
\!\!\!&+&\!\!\!\frac{4931}{8372} \tilde{\lambda}^4 +\frac{45}{2093})\r^{18}+\frac{4425}{2876} ( \tilde{\lambda}^{16}+\frac{38731}{17700} \tilde{\lambda}^{12} +\frac{4828817}{424800} \tilde{\lambda}^8 +\frac{2516369}{238950} \tilde{\lambda}^4-\frac{647}{4425}) \tilde{\lambda}^2 \r^{16}\cr &&\cr
\!\!\!&+&\!\!\!\frac{2790}{719} (\tilde{\lambda}^{12} +\frac{409321}{66960} \tilde{\lambda}^8 +\frac{18377303}{4821120} \tilde{\lambda}^4 +\frac{11389}{14880}
) \tilde{\lambda}^4 \r^{14}+\frac{341}{719}  (\tilde{\lambda}^{12}+\frac{9093}{352} \tilde{\lambda}^8 +\frac{26462905}{589248} \tilde{\lambda}^4\cr &&\cr 
\!\!\!&+&\!\!\!\frac{106413}{10912}) \tilde{\lambda}^6 \r^{12}
+\frac{62087}{17256} (\tilde{\lambda}^8 +\frac{10224109}{2235132} \tilde{\lambda}^4 +\frac{7861183}{1986784})\tilde{\lambda}^8 \r^{10}
+\frac{1301}{2876} (\tilde{\lambda}^8 +\frac{562726}{35127} \tilde{\lambda}^4 \cr &&\cr
\!\!\!&+&\!\!\!\frac{107584111}{2248128}) \tilde{\lambda}^{10} \r^8+(\frac{266509}{155304}  \tilde{\lambda}^{4}+\frac{11499791}{621216} ) \tilde{\lambda}^{12} \r^6
+(\frac{4367}{25884}  \tilde{\lambda}^{4}+\frac{213407}{23008} )  \tilde{\lambda}^{14}\r^4\cr &&\cr
&+&\frac{795431}{310608} \tilde{\lambda}^{16} \r^2 +\frac{93911}{310608} \tilde{\lambda}^{18} \bigg)\,.\nonumber
\eea
\bea
F_3\!\!\!&=&\!\!\!
-\frac{2228}{3}\bigg(\tilde{\lambda}^4\r^{28}+(\frac{1751323}{481248}\tilde{\lambda}^4-\frac{10}{557})
\tilde{\lambda}^2\r^{26}+(\frac{22017197}{3849984}\tilde{\lambda}^4+\frac{403}{1671})\tilde{\lambda}^4\r^{24}+(\frac{988465}{160416}\tilde{\lambda}^{8} \cr &&\cr
\!\!\!&+&\!\!\!\frac{92237}{80208}\tilde{\lambda}^4-\frac{1231}{6684})\tilde{\lambda}^2\r^{22}
+\frac{80353}{13368}(\tilde{\lambda}^8+\frac{1926317}{5785416}\tilde{\lambda}^4-\frac{20513}{160706})\tilde{\lambda}^4\r^{20}+\frac{8617}{2228}(\tilde{\lambda}^{12}\cr &&\cr
\!\!\!&+&\!\!\!\frac{2031685}{1861272}\tilde{\lambda}^8-\frac{91487}{155106}\tilde{\lambda}^4+\frac{2362}{25851})\tilde{\lambda}^2\r^{18}+\frac{1587}{2228}(\tilde{\lambda}^{16}+\frac{55925}{6348}\tilde{\lambda}^{12}
-\frac{76121}{16928}\tilde{\lambda}^8\cr &&\cr
\!\!\!&+&\!\!\!\frac{25091}{9522}\tilde{\lambda}^4-\frac{270}{529})\r^{16}+\frac{5161}{1671}(\tilde{\lambda}^{12}-\frac{55141}{495456}\tilde{\lambda}^8+\frac{663257}{743184}\tilde{\lambda}^4-\frac{7575}{20644}
)\tilde{\lambda}^2\r^{14}+\frac{463}{557}(\tilde{\lambda}^{12}\cr &&\cr
\!\!\!&+&\!\!\!\frac{4315}{44448}\tilde{\lambda}^8+\frac{3655517}{800064}\tilde{\lambda}^4-\frac{25291}{7408})\tilde{\lambda}^4\r^{12}+\frac{5821}{13368}(\tilde{\lambda}^8+\frac{125535}{23284}\tilde{\lambda}^4-\frac{223471}{23284})\tilde{\lambda}^6\r^{10}\cr &&\cr
\!\!\!&+&\!\!\!\frac{347}{2228}(\tilde{\lambda}^8+\frac{99827}{18738}\tilde{\lambda}^4-\frac{16771699}{599616})
\tilde{\lambda}^8\r^8+\frac{18701}{120312}(\tilde{\lambda}^4-\frac{376574}{18701})\tilde{\lambda}^{10}\r^6+(\frac{301}{20052}\tilde{\lambda}^{4}\cr &&\cr
\!\!\!&-&\!\!\!\frac{78325}{53472})\tilde{\lambda}^{12}\r^4
-\frac{24203}{60156}\tilde{\lambda}^{14}\r^2-\frac{11795}{240624}\tilde{\lambda}^{16}\bigg)\,.\nonumber
\eea
\bea
F_4\!\!\!&=&\!\!\!
324 \bigg( {\tilde{\lambda}}^{4}\r^{28}+{\frac {1027}{324}}{\tilde{\lambda}}^{6}\r^{26}
+ ({\frac {1893883}{559872}}{\tilde{\lambda}}^{4}+ {\frac {17}{27}} ){\tilde{\lambda}}^{4}\r^{24}
+ ( {\frac {3103}{1458}}{\tilde{\lambda}}^{8}+{\frac {2747}{1296}}{\tilde{\lambda}}^{4}-{\frac {20}{27}} ){\tilde{\lambda}}^{2}\r^{22}\cr &&\cr
\!\!\!&+&\!\!\!{\frac {2479}{648}} ( {\tilde{\lambda}}^{8}+{\frac {121927}{535464}}{\tilde{\lambda}}
^{4}-{\frac {7135}{14874}} ){\tilde{\lambda}}^{4} \r^{20}+{\frac {1775}{324}} ( {\tilde{\lambda}}^{12}-{\frac {6347}{21300}}{\tilde{\lambda}}^{4}
-{\frac {125927}{383400}}{\tilde{\lambda}}^{8}-{\frac {28}{355}} ) {\tilde{\lambda}}^{2}\r^{18}\cr &&\cr
\!\!\!&+&\!\!\!{\frac {295}{108}}( {\tilde{\lambda}}^{16}+{\frac {401}{1180}}{\tilde{\lambda}}^{12}-{\frac {6217}{50976}
}{\tilde{\lambda}}^{8}-{\frac {4199}{5310}}{\tilde{\lambda}}^{4}+{\frac {6}{59}} )  \r^{16}+
{\frac {1139}{324}}( {\tilde{\lambda}}^{12}+{\frac {2645}{4824}}{\tilde{\lambda}}^{8}-{\frac {14053}{13668}}{\tilde{\lambda}}^{4} \cr &&\cr
\!\!\!&+&\!\!\!{\frac {300}{1139}}){\tilde{\lambda}}^{2}  \r^{14}+{\frac {61}{81}} ( {\tilde{\lambda}}^{12}+{\frac{104215}{17568}}{\tilde{\lambda}}^{8}-{\frac {520673}{105408}}{\tilde{\lambda}}^{4}+{\frac {2819}{976}} ) {\tilde{\lambda}}^{4}\r^{12}+{\frac {1063}{486}} ( {\tilde{\lambda}}^{8}-{\frac {30295}{51024}}{\tilde{\lambda}}^{4}\cr &&\cr
\!\!\!&+&\!\!\!{\frac {11235}{8504}} ){\tilde{\lambda}}^{6} \r^{10}+{\frac {137}{324}} ({\tilde{\lambda}}^{8}-{\frac {211}{2466}}{\tilde{\lambda}}^{4} +{\frac {1529707}{236736}}){\tilde{\lambda}}^{8} \r^{8}+{\frac {1921}{17496}}( {\tilde{\lambda}}^{4}+{\frac {129061}{7684}} ){\tilde{\lambda}}^{10} \r^{6}\cr &&\cr
\!\!\!&+&\!\!\!{\frac {79}{2916}} ({\tilde{\lambda}}^{4} +{\frac {19495}{632}} ) {\tilde{\lambda}}^{12}\r^{4}+{\frac {3937}{17496}}{\tilde{\lambda}}^{14}\r^{2}+{\frac {955}{34992}}{\tilde{\lambda}}^{16}\bigg)\,.\nonumber 
\eea
\section*{Appendix B}
\bea
A&=& 
2 ( {r}^{4}-{{r_0}}^{4} ) r+{\frac {{{r_0}}^{2}\gamma}{12600{r}^{13}}} \bigg( 12073824{\lambda}^{2}{r}^{18}
+21181440{\lambda}^{2}{r}^{18}\ln  ( 1+{\frac {{{r_0}}^{2}}{{r}^{2}}} ) \cr &&\cr
&-& 21181440{\lambda}^{2}{{r_0}}^{2}{r}^{16}-21181440{r}^{14}{\lambda}^{2}{{r_0}}^{4}\ln  ( 2 ) 
-2023875{{r_0}}^{2}{r}^{14}-50400{r}^{14}{\lambda}^{2}{{r_0}}^{2}{\omega}^{2}\cr &&\cr 
&+&27404704{r}^{14}{\lambda}^{2}{{r_0}}^{4}-12932080{r}^{12}{\lambda}^{2}{{r_0}}^{6}
+39900{r}^{10}{\lambda}^{2}{{r_0}}^{6}{\omega}^{2}+307125{{r_0}}^{6}{r}^{10}\cr &&\cr
&-&34650{r}^{8}{{r_0}}^{6}{\omega}^{2}-3686088{r}^{8}{{r_0}}^{10}{\lambda}^{2}
+14700{r}^{6}{\lambda}^{2}{{r_0}}^{10}{\omega}^{2}+5748750{r}^{6}{{r_0}}^{10}\cr &&\cr
&-&19020720{r}^{4}{{r_0}}^{14}{\lambda}^{2}-4032000{r}^{2}{{r_0}}^{14}+17341800{{r_0}}^{18}{\lambda}^{2} \bigg) \cr &&\cr
B&=&
\frac32 ( {r}^{4}-{{r_0}}^{4} ) r+{\frac { {{r_0}}^{2}\gamma}{16800{r}^{13}}}\bigg( 12073824{\lambda}^{2}{r}^{18}+21181440{\lambda}^{2}{r}^{18}
\ln  ( 1+{\frac {{{r_0}}^{2}}{{r}^{2}}} ) \cr &&\cr
&-&22970640{\lambda}^{2}{{r_0}}^{2}{r}^{16}-21181440{r}^{14}{\lambda}^{2}{{r_0}}^{4}\ln(2)
-2023875{{r_0}}^{2}{r}^{14}+201600{r}^{14}{\lambda}^{2}{{r_0}}^{2}{\omega}^{2}\cr &&\cr
&+&27404704{r}^{14}{\lambda}^{2}{{r_0}}^{4}-12570180{r}^{12}{\lambda}^{2}{{r_0}}^{6}
+336000{r}^{10}{\lambda}^{2}{{r_0}}^{6}{\omega}^{2}+838425{{r_0}}^{6}{r}^{10}\cr &&\cr
&-&201600{r}^{8}{{r_0}}^{6}{\omega}^{2}-5600588{r}^{8}{{r_0}}^{10}{\lambda}^{2}
+268800{r}^{6}{\lambda}^{2}{{r_0}}^{10}{\omega}^{2}+5710950{r}^{6}{{r_0}}^{10}\cr &&\cr
&-&19007420{r}^{4}{{r_0}}^{14}{\lambda}^{2}-4088700{r}^{2}{{r_0}}^{14}+17489500{{r_0}}
^{18}{\lambda}^{2}\bigg)\cr &&\cr
C&=&
\frac{4}{r^2}\bigg(3{r}^{6}+{\lambda}^{2}{{r_0}}^{4}{r}^{4}-{r}^{2}{{r_0}}^{4}-{\lambda}^{2}{{r_0}}^{8}\bigg)
+{\frac {{{r_0}}^{2}\gamma}{12600{r}^{14}}}\bigg( 2299500{r}^{14}{{r_0}}^{6}-37910250{r}^{10}{{r_0}}^{10}\cr &&\cr
&-&43218000{r}^{2}{{r_0}}^{18}+790650{r}^{8}{{r_0}}^{10}{\omega}^{2}+81742500{r}^{6}{{r_0}}^{14}
-546860235{r}^{4}{{r_0}}^{18}{\lambda}^{2}\cr &&\cr
&+&341269458{r}^{8}{{r_0}}^{14}{\lambda}^{2}-49327173{r}^{12}{{r_0}}^{10}{\lambda}^{2}
-50575808{r}^{14}{\lambda}^{2}{{r_0}}^{8}+72793560{r}^{16}{\lambda}^{2}{{r_0}}^{6}\cr &&\cr
&+&20495744{\lambda}^{2}{{r_0}}^{4}{r}^{18}-122473890{\lambda}^{2}{{r_0}}^{2}{r}^{20}
+72442944{\lambda}^{2}{r}^{22}+262235400{{r_0}}^{22}{\lambda}^{2}\cr &&\cr
&+&127088640{r}^{18}(r^4-r_0^4){\lambda}^{2}\ln(1+{\frac {{{r_0}}^{2}}{{r}^{2}}} ) 
-42362880{r}^{14}{{r_0}}^{4}(r^4-r_0^4){\lambda}^{2}\ln ( 2 )\cr &&\cr
&-&22050{r}^{12}{{r_0}}^{6}{\omega}^{2}-226800{r}^{18}{\lambda}^{2}{{r_0}}^{2}{\omega}^{2}
-2913750{r}^{18}{{r_0}}^{2}-497700{r}^{14}{\lambda}^{2}{{r_0}}^{6}{\omega}^{2}\cr &&\cr
&-&1421700{r}^{10}{\lambda}^{2}{{r_0}}^{10}{\omega}^{2}
-1465800{r}^{6}{\lambda}^{2}{{r_0}}^{14}{\omega}^{2} \bigg)\frac{1}{( {r}^{4}-{{r_0}}^{4} )} \cr &&\cr
D&=&
-\frac {1}{{ 2( {r}^{4}-{{r_0}}^{4} ) {r}^{3}}} \bigg(( -8+{\lambda}^{2}{\omega}^{2} ) {r}^{10}- ( 8{\lambda}^{2}{{r_0}}^{4}+{\omega}^{2} ) {r}^{8}+ ( 8-{\lambda}^{2}{\omega}^{2} ) {{r_0}}^{4}{r}^{6}+8{\lambda}^{2}{{r_0}}^{12}\bigg)\cr &&\cr
&+&{\frac {\gamma}{100800{r}^{15} ( {r}^{4}-{{r_0}}^{4} ) ^{2}{{r_0}}^{2}}} \bigg( 120738240{\lambda}^{2}{r}^{18} ( -{\frac {32782}{17967}}{{r_0}}^{4}{\omega}^{2}{r}^{6}-{\frac {100864}{17967}}{{r_0}}^{8}{r}^{4}+{\omega}^{2}{r}^{10}\cr &&\cr
&+&{\frac {2837}{5989}}{{r_0}}^{8}{\omega}^{2}{r}^{2}+{\frac {50432}{17967}}{{r_0}}^{4}{r}^{8}
+{\frac {50432}{17967}}{{r_0}}^{12} ) \ln  ( 1+{\frac {{{r_0}}^{2}}{{r}^{2}}} ) 
+42362880{\lambda}^{2}{{r_0}}^{8}{r}^{20}\ln  ( 2 ) {\omega}^{2}\cr &&\cr
&+&193181184{\lambda}^{2}{{r_0}}^{2}(-\frac58{\omega}^{2}+{{r_0}}^{2}){r}^{26}
+(-301985040{\lambda}^{2}{{r_0}}^{6}+72442944{\lambda}^{2}{{r_0}}^{4}{\omega}^{2} ) {r}^{24}\cr &&\cr
&+&( -216910848{\lambda}^{2}{{r_0}}^{8}+180332460{\lambda}^{2}{{r_0}}^{6}{\omega}^{2} ) {r}^{22}+ (588652400{\lambda}^{2}{{r_0}}^{10}\cr &&\cr &-&151400000{\lambda}^{2}{{r_0}}^{8}{\omega}^{2}+( 4473000{\omega}^{2}-756000{\lambda}^{2}{\omega}^{4} ) {{r_0}}^{6} ) {r}^{20} 
+ ( 23729664{\lambda}^{2}{{r_0}}^{12}\cr &&\cr
&+& ( -4914000+9837212{\lambda}^{2}{\omega}^{2} ) {{r_0}}^{10} ) {r}^{18}+( 1366241912{\lambda}^{2}{{r_0}}^{14}+ ( -590625{\omega}^{2} \cr &&\cr &-&1314600{\lambda}^{2}{\omega}^{4} ) {{r_0}}^{10} ) {r}^{16}+(( -45108000+9495348{\lambda}^{2}{\omega}^{2} ) {{r_0}}^{14} +466200{{r_0}}^{10}{\omega}^{4} ) {r}^{14}\cr &&\cr
&+& ( -7974974584{\lambda}^{2}{{r_0}}^{18}+ ( -331800{\lambda}^{2}{\omega}^{4}-15600375{\omega}^{2} ) {{r_0}}^{14} ) {r}^{12}+( 48816276{\lambda}^{2}{\omega}^{2}\cr &&\cr
&+&159390000 ) {{r_0}}^{18}{r}^{10}
+ ( 6879600{{r_0}}^{18}{\omega}^{2}+12897439512{\lambda}^{2}{{r_0}}^{22} ) {r}^{8}+ ( -163800000\cr &&\cr
&-&29432400{\lambda}^{2}{\omega}^{2} ) {{r_0}}^{22}{r}^{6}
-8635726200{\lambda}^{2}{{r_0}}^{26}{r}^{4}+54432000{{r_0}}^{26}{r}^{2}+2060352000{\lambda}^{2}{{r_0}}^{30} \bigg) \cr &&\cr
E&=&
-\frac{\gamma {{r_0}}^{4}}{24r^{11}}  \bigg( 204{\lambda}^{2}{r}^{16}-317{\lambda}^{2}{{r_0}}^{4}{r}^{12}
-111{{r_0}}^{4}{r}^{10}+299{\lambda}^{2}{{r_0}}^{8}{r}^{8}+222{{r_0}}^{8}{r}^{6}\cr &&\cr
&-& 463{\lambda}^{2}{{r_0}}^{12}{r}^{4}-111{{r_0}}^{12}{r}^{2}+277{\lambda}^{2}{{r_0}}^{16} \bigg)\cr &&\cr
F&=&
-\frac{\gamma{{r_0}}^{4}}{12r^{12}}  \bigg( 708{\lambda}^{2}{r}^{16}-753{\lambda}^{2}{{r_0}}^{4}{r}^{12}
-267{{r_0}}^{4}{r}^{10}+805{\lambda}^{2}{{r_0}}^{8}{r}^{8}+378{{r_0}}^{8}{r}^{6}\cr &&\cr
&-&1047{\lambda}^{2}{{r_0}}^{12}{r}^{4}-111{{r_0}}^{12}{r}^{2}+287{\lambda}^{2}{{r_0}}^{16}\bigg) \nonumber
\eea
\section*{Appendix C}
Introducing the $\omega=2r_0\wn$ the coefficients in equation of motion are
\bea
C_1&=&-17( {\r}^{4}-1) ^{4}{\r}^{3}\bigg( \tilde{\lambda}^{2}(\r^8
+{\frac {91}{204}}{\r}^{4}+{\frac {277}{204}}) -{\frac {37}{68}}{\r}^{2}\bigg)\cr &&\cr
C_2&=&-\frac16(\r^4-1)^3{\r}^{2}\bigg(1020{\tilde\lambda}^{2}{\r}^{12}
+703{\tilde\lambda}^{2}{\r}^{8}+111{\r}^{6}-194{\tilde\lambda}^{2}{\r}^{4}-999{\r}^{2}+3047{\tilde\lambda}^{2}\bigg)\cr &&\cr
C_3&=&-{\frac {25216}{15}} ( {\r}^{4}-1 )^{2}\r \bigg(-\frac12{\r}^{18}{\tilde\lambda}^{2}\ln( 1+\frac{1}{\r^2}) +\frac12{\r}^{14}{\tilde\lambda}^{2}\ln(2)-{\frac{17967}{63040}}{\r}^{18}{\tilde\lambda}^{2}\cr &&\cr 
&+&{
\frac {18053}{25216}}{\tilde{\lambda}}^{2}{\r}^{16}+ ( {\frac {19275}{403456}}
+{\frac {15}{197}}{\tilde{\lambda}}^{2} ( -{\frac {856397}{100800}}+{\wn}^{2} )){\r}^{14}+{\frac {100127}{302592}}{\tilde{\lambda}}^{2}{\r}^{12}\cr &&\cr
&+&{\frac {505}{6304}}( {\frac {3699}{6464}}+{\tilde{\lambda}}^{2}{\wn}^{2} ) {\r}^{10}+( -{\frac {15293}{504320}}{\tilde{\lambda}}^{2}-{\frac {555}{12608}}{\wn}^{2} ) {\r}^{8}
+{\frac {445}{6304}} ( {\frac {3201}{2848}}\cr &&\cr &+&{\tilde{\lambda}}^{2}{\wn}^{2} ) {\r}^{6}-{\frac {296643}{706048}}{\r}^{4}{\tilde{\lambda}}^{2}-{\frac {35205}{100864}}{\r}^{2}
+{\frac {125365}{100864}}{\tilde{\lambda}}^{2} \bigg) \cr &&\cr
C_4&=&-{\frac {25216}{3}} ( {\r}^{4}-1 ) ^{2} \bigg( -\frac12{\r}^{20}{\tilde{\lambda}}^{2} 
( 1+\frac{1}{{\r}^{2}} ) \ln  ( 1+{\frac {1}{{\r}^{2}}} ) +\frac{1}{10}{\r}^{16}{\tilde{\lambda}}^{2} 
( 1+\frac{1}{{\r}^{2}}  ) \ln  ( 2 )\cr &&\cr
&-&{\frac {17967}{63040}}{\r}^{20}{\tilde{\lambda}}^{2}+{\frac {28141}{126080}}{\r}^{18}{\tilde{\lambda}}^{2}
+{\frac {3}{197}} ( {\frac {1285}{2048}}+{\tilde{\lambda}}^{2} ( {\wn}^{2}+{\frac {2367401}{201600}} )) {\r}^{16}\cr &&\cr
&+&{\frac {3}{197}}({\frac {1285}{2048}}+{\tilde{\lambda}}^{2} ( {\wn}^{2}-{\frac {10355621}{806400}} )  ) {\r}^{14}
-{\frac {303}{6304}}( {\frac {4587}{6464}}+{\tilde{\lambda}}^{2} ( {\frac {100127}{72720}}\cr &&\cr &+&{\wn}^{2} ) ) {\r}^{12}-( {\frac {303}{6304}}( {\wn}^{2}-{\frac {104993}{24240}} ) {\tilde{\lambda}}^{2}
-{\frac {555}{12608}}{\wn}^{2}+{\frac {13761}{403456}} ) {\r}^{10}\cr &&\cr
&-&{\frac {623}{6304}} ( {\tilde{\lambda}}^{2} ( {\wn}^{2}-{\frac {14999}{7120}} ) 
-{\frac {555}{1246}}{\wn}^{2}-{\frac {10119}{2848}} ) {\r}^{8}-{\frac {623}{6304}} ( -{\frac {10119}{2848}}\cr &&\cr 
&+&{\tilde{\lambda}}^{2} ( {\frac {5497533}{348880}}
+{\wn}^{2} )) {\r}^{6}+ ( -{\frac {32439}{100864}}-{\frac {5497533}{3530240}}{\tilde{\lambda}}^{2}) {\r}^{4}+( {\frac {149383}{100864}}{\tilde{\lambda}}^{2}\cr &&\cr &-&{\frac {32439}{100864}} ) {\r}^{2}
+{\frac {149383}{100864}}{\tilde{\lambda}}^{2}\bigg) \cr &&\cr
C_5&=&-{\r}^{3}\bigg( -{\frac {47912}{5}} ( {\frac {2837}{5989}}
-{\frac {32782}{17967}}{\r}^{4}+{\r}^{8} ) {\tilde{\lambda}}^{2}{\wn}^{2}{\r}^{16}\ln  ( 1+{\frac {1}{{\r}^{2}}} )+{\frac {47912}{5}} (\cr &&\cr
&-&{\frac {6304}{17967}}{\r}^{16}{\tilde{\lambda}}^{2}{\wn}^{2}\ln  ( 2 ) +{\r}^{22}{\tilde{\lambda}}^{2}{\wn}^{2}-\frac35{\r}^{20}{\tilde{\lambda}}^{2}{\wn}^{2}-{\frac {143121}{95824}}{\r}^{18}{\tilde{\lambda}}^{2}{\wn}^{2}\cr &&\cr
&+&{\frac {150}{5989}} ( {\tilde{\lambda}}^{2} ( {\wn}^{4}+{\frac {18925}{378}}{\wn}^{2}
+{\frac {13}{20}} )-{\frac {71}{48}}{\wn}^{2} ) {\r}^{16}-{\frac {5729}{4312080}}{\r}^{14}{\tilde{\lambda}}^{2}{\wn}^{2}\cr &&\cr
&+&{\frac {1565}{35934}} ( {\tilde{\lambda}}^{2} ( -{\frac {312}{313}}+{\wn}^{4} ) +{\frac {10053}{10016}}{\wn}^{2} ) {\r}^{12}
+ ( {\frac {1014907}{3353840}}{\tilde{\lambda}}^{2}{\wn}^{2}\cr &&\cr &-&{\frac {185}{11978}}{\wn}^{4} ) {\r}^{10}+{\frac {395}{35934}}( {\tilde{\lambda}}^{2} ( -{\frac {62}{79}}+{\wn}^{4} )
-{\frac {46701}{2528}}{\wn}^{2} ) {\r}^{8}-{\frac {1675}{35934}}{\tilde{\lambda}}^{2}\cr &&\cr &+&{\frac {96259}{3353840}}{\r}^{6}{\tilde{\lambda}}^{2}{\wn}^{2}+( {\frac {3225}{23956}}{\wn}^{2}+{\frac {1480}{17967}}{\tilde{\lambda}}^{2} ) {\r}^{4}
-{\frac {86525}{503076}}{\r}^{2}{\tilde{\lambda}}^{2}{\wn}^{2} ) \bigg) \nonumber
\eea



\end{document}